\documentclass[aps,notitlepage,superscriptaddress,pre]{revtex4-2}
\usepackage{ucs}
\usepackage[utf8x]{inputenc}
\usepackage{amsmath}
\usepackage{amsfonts}
\usepackage{amssymb}
\usepackage{amsthm}
\usepackage{fontenc}
\usepackage[dvips,pdftex]{graphicx}
\usepackage{setspace}
\usepackage{color}
\usepackage[dvipsnames]{xcolor}
\usepackage{graphicx}

\usepackage{hyperref}

\usepackage{multirow}
\usepackage{array}
\usepackage{caption}
\usepackage{subcaption}
\newcommand{\blu}[1]{#1}
\newcommand{\dt}{\,\text{d}}

\newcommand{\pd}{\partial}
\newcommand{\Tr}{\text{Tr}}
\newcommand{\unity}{\mathbb{I}}

\graphicspath{{pics/}}
 \DeclareGraphicsExtensions{.pdf,.eps,.jpg}

\hypersetup{linktocpage}

\begin{document}
\begin{abstract}
Using a statistical-mechanics approach, we study the effects of  geometry and self-avoidance on the ordering of slender filaments inside non-isotropic containers, considering cortical microtubules in plant cells, and packing of genetic material inside viral capsids as concrete examples. Within a mean-field  approximation, we show analytically how the shape of the container, together with self-avoidance, affects the ordering of the stiff rods.  We find that the strength of the self-avoiding interaction plays a significant role in the preferred packing orientation, leading to a first-order transition  for oblate cells, where the preferred orientation changes from azimuthal, along the equator, to a polar one,  when self-avoidance is strong enough.  While for prolate spheroids the ground state is always a polar-like order, strong self-avoidance results with a deep meta-stable state along the equator. We compute the critical surface describing the transition between  azimuthal and polar ordering in the three dimensional parameter space (persistence length, eccentricity, and self-avoidance) and show that the critical behavior of this system is in fact related to the butterfly catastrophe model. We calculate the pressure and shear stress applied by the filament on the surface, and the injection force needed to be applied on the filament in order to insert it into the volume.  We compare these results to the pure mechanical study where self-avoidance is ignored, and discuss similarities and differences. 
\end{abstract} 
\title{}
\title{Effects of self-avoidance on the packing of stiff rods on ellipsoids}
\author{Doron \surname{Grossman}}
\email[]{doron.grossman@ladhyx.polytechnique.fr}

\affiliation{LadHyX, CNRS, Ecole polytechnique, Institut Polytechnique de Paris, 91128 Palaiseau Cedex, France}
\author{Eytan \surname{Katzav}}
\email[]{eytan.katzav@mail.huji.ac.il }
\affiliation{Racah Institute of Physics, Hebrew University, Jerusalem 9190401, Israel}	
\date{\today}
\maketitle

\section{Introduction}
The manner by which filaments pack inside a given shape is an important problem in many fields. From the ordering of micro-tubules inside plant cells \cite{uyttewaal2012mechanical,burk2002alteration,stoop2011morphogenesis}, via packing of genetic material inside viral capsids\cite{marenduzzo2010biopolymer,katzav2006statistical,boue2007folding,purohit2003mechanics,marenduzzo2008computer,marenduzzo2009statistics,ali2006polymer}, through  spooling of filaments and biopolymers  \cite{vetter2013finite,vetter2015growth,pineirua2013spooling,elettro2015elastocapillary,elettro2017drop,elettro2017elastocapillary,elsner2012spatiotemporal}.  
Most approaches to such problems are numerical, and regard highly symmetric volumes, as these are hard problems that involve elasticity, entropy, self-avoidance, and global (shape) constraints. As such these systems are inherently frustrated, i.e. exhibit residual stresses, in the absence of any external forcing, that arise directly from the filaments being limited onto a confining manifold \cite{Efrati2009,Aharoni2014geometry,Armon2011}. As a result of this, a complex configuration space is expected to arise, affecting the statistics and mechanics of such system  \cite{Grossman2016,Grossman2018shape}.

  The aim of this paper is an analytical study of systems which are not highly symmetric, using statistical mechanics. As concrete examples,  we consider cortical microtubules in plant cells, which play a significant role in cell shape regulation, and packing of RNA inside viral capsids.  This paper is, in some ways, an expansion of previous work  \citep{grossman2021packing}, using different methods so as to incorporate the effects of noise (finite persistence length) and self-avoidance.

In the  context of plant tissue, cellulose fibrils are deposited along the orientation of microtubules (MT) on the cells' membrane \cite{burk2002alteration}. These fibrils, being the main mechanical load carriers, set the mechanical properties of the cell wall. Thus, any anisotropy in their orientation and deposition, results in an anisotropic response of the cell to internal and external loads. This serves as the underlying mechanism for cell shape regulation.  As the MT orientation decides the orientation of cellulose fibrils, the question becomes that of understanding what sets the MT orientation  at the single cell level. It was suggested that MT sense directly the mechanical stress on the cell wall \cite{uyttewaal2012mechanical}. On the other hand, the mechanism by which the stress on the cell wall affects the membrane \cite{fisher1998extending,somerville2004toward} is not completely clear. Application of stress on the wall seems to affect the mechanical response only after a delay of several hours \cite{sahaf2016rheology}, suggesting an integral (thus time delayed) mechanism, which is common in other growth regulation mechanisms {\cite{oh2014cell,chaiwanon2016information}}. Alternatively,  the cell's shape anisotropy is suggested  to play a dominant role in the ordering of MTs. The last two approaches are closely related, as an integral over the strain results with the shape observed. Hence, we aim to identify the leading effects of cell shape on the disordered-aligned transition of MT.  As MTs cannot cross each other and tend to align together, this interaction cannot be disregarded. Thus, we study the manner by which cell anisotropy, filament stiffness, and self-avoidance affect the order-disorder transitions.

In the context of packing of genetic material inside viral capsids, this paper aims to elucidate further on the effect of geometry on the packing, which is essential in understanding the ejection of genetic material during infection \cite{de2005dna,purohit2003force,purohit2003mechanics,purohit2005forces,kindt2001dna,smith2001bacteriophage}. Though it is likely to play an important role, in this paper we do not discus the effects of electrostatic interaction \cite{fathizadeh2013confinement,nikoubashman2017semiflexible,milchev2018adsorption,netz2003neutral,borukhov1997steric,abrashkin2007dipolar,slosar2006connected,van2005electrostatics,ghosh2002phase,carri1999attractive,angelescu2008viruses}, and we leave this to a later work. Here we will focus on elucidating the effects of the container shape, and of self-avoidance.

In this work, we consider the filament to be stiff and narrow so that it effectively adheres to the surface of an ellipsoid. Yet, it may freely choose an orientation on the surface. This constraint is somewhat relaxed as we will be working within a mean-field approximation, {following the approach developed in Refs. \cite{gupta1993mean,koga2004jamming} so that the filament, conforms to the surface only on average. We are thus ignoring both temporal fluctuations and spatial variations of the different fields}. As the surface is not uniform nor highly symmetric, such an approximation is valid for large persistence lengths, namely greater than some typical scale of the ellipsoid{, and in cases where the ellipsoidal containers are not too eccentric nor too flat. The main technical challenge that our mean-field approach has to overcome is related to self-avoidance which is known to be notoriously difficult. At the same time, mean-field methods developed by Edwards in the context of self-avoiding polymers \cite{edwards1965statistical} provide excellent qualitative, and often also quantitative, description of such systems. At any rate, the huge experience with mean-field methods across many branches of statistical physics showed that it is always worth exploring, and provides at least a useful global understanding of the phase diagram of the system. A classical example is that of the Ising model, where mean-field theories reproduce correctly the qualitative phenomenology, including the phase diagram and the existence of critical exponents, yet fail to provide the exact values of the exponents.}


\section{Derivation}

Following Refs. \cite{katzav2006statistical,boue2007folding}, we begin with the energy of a long, self-avoiding, rod, of length $L$, on an ellipsoid:
\begin{align} \label{eq: energy}
	E&= \frac{\tilde{\epsilon}}{2} \int_0^L \dt s \mathbf{R}''^2 + \frac{\tilde{u}}{2} \int_0^L\int_0^L \dt s_1 \dt s_2 \left[\mathbf{R}'(s_1) \times \mathbf{R}'(s_2)\right]^2 \delta \left( \mathbf{R}(s_1)-\mathbf{R}(s_2) \right).
\end{align}
The first term on the right hand side is the elastic energy associated with bending the rod. The second term describes the self-avoiding interaction \cite{mehta1994granular,gupta1993mean,koga2004jamming,edwards1965statistical}. Here $\tilde{\epsilon}$ is the persistence length, $\tilde{u}$ measures the strength of the self-avoidance and is dimensionless, $\mathbf{R}(s)=\left(X(s),Y(s),Z(s)\right)$, is the configuration (or position) vector and is constrained to the surface of the ellipsoids, namely $\Sigma_1\left({X}^2 + Y^2\right) +\Sigma_3 Z^2 = \ell^2$ (where $\ell$ sets a length scale, which will later be related to the area of the ellipsoid) and is inextensible, which can be enforced by the choice 
\begin{align}
{\mathbf{R}'}^2 =1.
\end{align} 
In the former, $\Sigma_1$ and $\Sigma_3  $ are  some general (axi-symmetric) scalings of the axes.

The free energy, $W$, of the system is given using the following path integral:
\begin{align}\label{eq: Free energy}
	e^{-W} =  \int \mathcal{D}[{\bf R}]   \;\delta\left(\Sigma_1 X^2 +\Sigma_1 Y^2+\Sigma_3 Z^2 -\ell^2\right) \;\delta\left(\mathbf{R'}^2-1\right)  e^{-\frac{\tilde{\epsilon}}{2} \int \dt s \mathbf{R}''^2 - \frac{\tilde{u}}{2} \iint \dt s_1 \dt s_2 \left[\mathbf{R}'(s_1) \times \mathbf{R}'(s_2)\right]^2 \delta (\mathbf{R}(s_1)-\mathbf{R}(s_2))} .
\end{align}

\noindent
In order to advance, we introduce the orientation tensor {${S}_{ij}$}
\begin{align}\label{eq: orientation_tensor_1}
	 S_{ij}(\mathbf{r}) &= \int \dt s R_i' R_j' ~\delta(\mathbf{r}- \mathbf{R}(s)).
\end{align}
{The orientation tensor quantifies the alignment of the filament on a specific position on the surface of the ellipsoid}.
We enforce this definition into the path integral by {introducing a $\delta$-function into the path integral and further exponentiating it using} the auxiliary field $\psi_{ij}$
\begin{align}
	{1 =} \int \mathcal{D}[{S}] \, \delta\left[ {S}_{ij} -\int \dt s R'_i R'_j \delta\left(\mathbf{r}-\mathbf{R}(s)\right)\right] &= N_0 \int \mathcal{D}[S] \int \mathcal{D}[\psi] e^{i \int_\Omega \dt^2 r \psi_{ij}(\mathbf{r}) {S}_{ij}(\mathbf{r}) - i \int \dt s R'_i \psi_{ij}(\mathbf{R}(s))R'_j},
\end{align}

\noindent
where we assume a summation convention, so that repeating indexes are summed over, and $N_0$ is an unimportant normalization constant. Note that this integral is {simply equal to one}. Enforcing inextensibility and constraining to the surface of the ellipsoid is done in a similar manner, by introducing the auxiliary fields $\lambda$ and $\chi$ 
\begin{align}\label{eq: }
	\delta\left(\Sigma_1 (X^2 +Y^2)+\Sigma_3 {Z^2} -\ell^2\right) &= N_1 \int \mathcal{D}[\chi]e^{-i \int \dt s \chi(s) \left(\Sigma_1 X^2+\Sigma_1 Y^2+\Sigma_3 Z^2- \ell^2\right)} , \\
	\delta(\mathbf{R}'^2-1)&= N_2 \int \mathcal{D} [\lambda] e^{-i \int \dt s \lambda(s)\left(\mathbf{R}'^2- 1\right)},
\end{align}

\noindent
where $N_i$'s are normalization constants, which do not contribute to the final result. Inserting these definitions and noting that the self-avoidance term can be expressed in terms of the orientation tensor (\ref{eq: orientation_tensor_1})

\begin{align}
	&\iint \dt s_1 \dt s_2 \left[\mathbf{R}'(s_1) \times \mathbf{R}'(s_2)\right]^2 \delta \left(\mathbf{R}(s_1)-\mathbf{R}(s_2) \right) \\ \nonumber &=  \iint \dt s_1 \dt s_2 R'
	_i(s_1)R'_k(s_1) R'_j(s_2) R'_l(s_2) \left(\delta_{ik}\delta_{jl}-\delta_{il}\delta_{jk}\right) \delta\left(\mathbf{R}(s_1) -\mathbf{R}(s_2) \right) \\ \nonumber &= \int \dt r \left(\delta_{ik}\delta_{jl}-\delta_{il}\delta_{jk}\right) \left\{\int \dt s_1  \left[ R'
	_i(s_1)R'_k(s_1) \delta (\mathbf{r} - \mathbf{R}(s_1)) \right]  \times \int \dt s_2 \left[ R'
	_j(s_2)R'_l(s_2) \delta (\mathbf{r} - \mathbf{R}(s_2)) \right] \right\} \\ \nonumber
	& = \int \dt r \left(\delta_{ik}\delta_{jl}-\delta_{il}\delta_{jk}\right) \left\{S_{ik} S_{jl} \right\} = \int \dt r \left(S_{ii}S_{jj}-S_{ij}S_{ij}\right), 
\end{align}

\noindent
we get

\begin{align}
	e^{-W} &= N \int \mathcal{D}[{\bf R}]\int {D}[\lambda] \int {D}[\chi]\int {D}[\psi] \int {D}[{S}]  e^{- \int \dt s \left\{ \frac{\tilde{\epsilon}}{2} \mathbf{R}''^2 + i \lambda \mathbf{R}'^2 + i R'_i \psi_{ij} R'_j + i \chi \left[\Sigma_1(X^2+Y^2)+\Sigma_3 Z^2- \ell^2\right]\right\}} \\ \nonumber
	& \times e^{i \int \dt s \lambda + i \int \dt s \chi \ell^2 + i\int \dt^2 r \psi_{ij}S_{ij}} e^{-\frac{u}{2}\int \dt^2 r \left({S}_{ii}{S}_{j}- {S}_{ij}{S}_{ij}\right)},
\end{align}
and $N= N_0 \times N_1 \times N_2$.

We now reformulate the constraint onto an ellipsoid surface using tensorial quantities

\begin{align}
	\Sigma_1(X^2+Y^2)+\Sigma_3 Z^2&= R_i \Sigma_{ij} R_j,
\end{align}

\noindent
where $\Sigma_{ij}$ is the form tensor

\begin{align}
	\Sigma&= \left(\begin{array}{ccc}
		\Sigma_1 & 0 & 0 \\
		0& \Sigma_2 & 0\\
		0& 0& \Sigma_3\\
	\end{array} \right).
\end{align}

\noindent 
{We assume}, without loss of generality, {that $\Sigma$} is diagonalized, and  the eigenvalues obey $\Sigma_1 = \Sigma_2 \neq \Sigma_3$. We may write

\begin{align}\label{eq: free energy full}
	e^{-W} &= N \int \mathcal{D}[{\bf R}]\int {D}[\chi]\int {D}[\lambda] \int {D}[\psi] \int {D}[S]  e^{- \int \dt s \left( \frac{\tilde{\epsilon}}{2}  \mathbf{R}''^2 + i \lambda \mathbf{R}'^2 + i R'_i \psi_{ij} R'_j + i \chi R_i \Sigma_{ij}R_j\right)} \\ \nonumber
	& \times e^{i \int \dt s \lambda + i \int \dt s \chi \ell^2 + i\int \dt^2 r \psi_{ij}S_{ij}} e^{-\frac{\tilde{u}}{2}\int \dt^2 r \left(S_{ii}S_{jj}- S_{ij}S_{ij}\right)}.
	\ \end{align}

In order to proceed, we now change to Fourier space ($\mathbf{R}(s) = \int \dt q \hat{\mathbf{R}}(q) e^{i q s} $) and calculate the field-dependent free energy, and work in the mean - field approach. \blu{In this approach, we assume the conjugate fields, $\lambda, \chi, \psi_{ij}$  are all constant. This translates to constraints being satisfied only on average rather than exactly. In the limit of a sphere ,however,  when $\Sigma_1=\Sigma_2=\Sigma_3$ this can be shown to be exact for large $\tilde{\epsilon}$, by virtue of homogeneity, and isotropy of the sphere. Therefore deviation from the mean field scale with $\delta$ to leading order.} Working in the mean field approximation where we assume the conjugate fields are constant in space and time ({ and hence also $q$-independent}) we find 

\begin{align}\label{eq: mean field calc}
	e^{-W} &= N \int \mathcal{D}[{\bf 	\hat{R}}(q)] \exp\left[{- \int \dt q \left( \frac{\tilde{\epsilon}}{2} q^4 \mathbf{\hat{R}(q)}^2 + i q^2 \lambda \mathbf{\hat{R}(q)}^2 + i q^2 \hat{R}_i \psi_{ij} \hat{R}_j + i \chi \hat{R}_i \Sigma_{ij}\hat{R}_j\right)}\right] \\ \nonumber
	& \times \exp\left[{i \int \dt s \lambda + i \int \dt \chi \ell^2 + i\int \dt^2 r \psi_{ij}{S}_{ij}-\frac{\tilde{u}}{2}\int \dt^2 r \left({S}_{ii}{S}_{j}- {S}_{ij}{S}_{ij}\right)}\right] \\
	&= N \left(\prod_{q} \sqrt{\frac{2\pi}{\det\left[\frac{\tilde{\epsilon}}{2}q^4 {\unity} + i(\lambda {\unity} +\psi) q^2 +i\Sigma \chi \right]}}\right)\times  e^{i \int \dt s \lambda + i \int \dt s \chi \ell^2 + i\int \dt^2 r \psi_{ij}{S}_{ij}-\frac{\tilde{u}}{2}\int \dt^2 r \left({S}_{ii}{S}_{jj}- {S}_{ij}{S}_{ij}\right)},
\end{align} 

\noindent
{where $\det[M]$ stands for the determinant of the matrix $M$, and $\unity$ is the unit matrix.} Hence, up to unimportant constants

\begin{align}
	W &= \frac{L}{2\pi} \int\limits_0^\infty \ln\left[\det(\frac{\tilde{\epsilon}}{2}q^4 {\unity}  + i(\lambda {\unity} +\psi) q^2 +i\Sigma \chi )\right] \dt q -i\lambda L -i\chi \ell^2 L -i A \psi_{ij}{S}_{ij} + \frac{\tilde{u}}{2}A \left({S}_{ii}{S}_{jj}-{S}_{ij}{S}_{ij}\right),
\end{align}

\noindent
where $A$ is the area of the {ellipsoid}. Since $\lambda,\chi,\psi_{ij}$ are {yet} undetermined we can absorb $i$ into their definition, and rewrite the free-energy in the form

\begin{align}\label{eq: mean field ener}
	W &= \frac{L}{2\pi} \int\limits_0^\infty \ln\left[\det(\frac{\tilde{\epsilon}}{2}q^4 {\unity} + (\lambda {\unity} +\psi) q^2 +\Sigma \chi )\right] \dt q -\lambda L -\chi \ell^2 L - A \psi_{ij}{S}_{ij} + \frac{\tilde{u}}{2}A \left({S}_{ii}{S}_{jj}-{S}_{ij}{S}_{ij}\right).
\end{align}

\noindent
{The terms next to the integral, namely}

\begin{align} \label{eq:s1}
{s = \lambda L + \chi \ell^2 L + A \psi_{ij}S_{ij} - \frac{\tilde{u}}{2}A \left(S_{ii}S_{jj}-S_{ij}S_{ij}\right)}
\end{align}

{are, in fact, the local entropy, $s$, associated with a given configuration.}

The mean field equations are then obtained by  equating the derivatives of $W$ to $0$, namely

\begin{align} \label{eq: MF_chi}
	\frac{\partial W}{\partial \chi}&=  \frac{L}{2\pi} \int\limits_0^\infty \Tr\left[\left(\frac{\tilde{\epsilon}}{2}q^4 {\unity} + (\lambda {\unity} +\psi)q^2+\Sigma \chi\right)^{-1}\Sigma\right] \dt q	-\ell^2 L =0 \\\label{eq: MF_lamb}
	\frac{\partial W}{\partial \lambda}&=  \frac{L}{2\pi} \int\limits_0^\infty  \Tr\left[\left(\frac{\tilde{\epsilon}}{2}q^4 {\unity} + (\lambda {\unity} +\psi)q^2+\Sigma \chi\right)^{-1} \right] q^2 \dt q- L =0 \\ \label{eq: MF_psi}
	\frac{\partial W}{\partial \psi_{ij}}&=  \frac{L}{2\pi} \int\limits_0^\infty  \left(\frac{\tilde{\epsilon}}{2}q^4 {\unity} + (\lambda {\unity} +\psi)q^2+\Sigma \chi\right)^{-1}_{ij} q^2 \dt q - A S_{ij} =0 \\ \label{eq: MF_sig}
	\frac{\partial W}{\partial S_{ij}}&=  - A \psi_{ij} + \tilde{u} A \left(\delta_{ij}\Tr[S]-S_{ij}\right)  =0 .
\end{align}

\noindent
{where $\Tr[M]$ is the trace of the matrix $M$.}

Equations \eqref{eq: MF_chi} - \eqref{eq: MF_sig} are integral matrix equations and are generally hard to solve. We therefore employ another simplification, and instead of using a local definition of {$S$} we replace it with a global average of it. \blu{Thus, $S$ will now be an average quantity, and we will lose a detailed configurational description.}

\begin{align}
{{S}^{\rm glob}_{ij}}= \frac{ \int {S_{ij}} \dt^2 r}{\int \dt^2 r} .
\end{align}

\noindent
Naturally, ${S}^{\rm glob}$ is diagonalized in the same coordinate system as $\Sigma$. Also, since 

\begin{align}
\Tr(R_i R_j) = 1 ,
\end{align}

\noindent
we may (abusing notation) write ${S}^{\rm glob}= c{S}^{\rm glob}$ where $c=\frac{L}{A}$ is the density of the filament per unit area on the spheroid. Since $\psi_{ij}$ is the auxiliary field to $S_{ij}$ (and as can be seen directly from  Eq.\eqref{eq: MF_sig}) it follows the same symmetries.  We can therefore write (omitting the "glob" superscript for simplicity )
\begin{align}
	\psi &= \left(\begin{array}{ccc}
		\psi_1 & 0 &0\\
		0& \psi_2 &0\\
		0& 0& \psi_3 \\
	\end{array}\right) ,\\
	{S}= {S}^{\rm glob} &= \left(\begin{array}{ccc}
		\sigma_1 & 0 &0\\
		0& \sigma_2 &0\\
		0& 0& \sigma_3 \\
	\end{array}\right)
\end{align}

\noindent
where $\sigma_1+\sigma_2+\sigma_3 =1$ leading to $\sigma_3= 1-\sigma_1 -\sigma_2$.
Thus,

\begin{align}
	\left(\frac{\tilde{\epsilon}}{2}q^4 {\unity} + (\lambda {\unity} +\psi)q^2+\Sigma \chi\right)^{-1} &=  \left(\begin{array}{ccc}
		\frac{1}{\tilde{\epsilon} q^4/2 + (\lambda +\psi_1)q^2+ \chi \Sigma_1} &0 &0 \\
		0 & \frac{1}{\tilde{\epsilon} q^4/2 + (\lambda +\psi_2)q^2+ \chi \Sigma_2} & 0 \\
		0 & 0& \frac{1}{\tilde{\epsilon} q^4/2 + (\lambda +\psi_3)q^2+ \chi\Sigma_3}\\
	\end{array}\right).
\end{align}

\noindent
Within this approximation, we can now write

\begin{align} \label{eq: Free energy MF}
	W& = 2\lambda L + 2 A c \psi_i\sigma_i + 4 \chi  \ell^2 L -\lambda L -\chi \ell^2 L -A c \psi_{i}\sigma_{i} + \frac{\tilde{u}}{2}A c^2 \left(1-\sigma_{i}\sigma_{i}\right) \\ \nonumber
	&= \lambda L +  L \psi_i\sigma_i + 3 \chi  \ell^2 L  + \frac{\tilde{u}}{2}c L \left(1-\sigma_{i}\sigma_{i}\right)
\end{align}

\noindent
{Going back to Eq. (\ref{eq:s1}), the contribution to the free energy, coming from the entropy, $s$, now reads}

\begin{align} \label{eq:s2}
{s =\lambda L +\chi \ell^2 L + A c \psi_{i}\sigma_{i} - \frac{\tilde{u}}{2}A c^2 \left(1-\sigma_{i}\sigma_{i}\right)}
\end{align}

{Looking for solutions with the ellipsoid symmetry imposed by the containers}, we search for solutions that obey $\sigma_1=\sigma_2 =\sigma$, and hence also $\sigma_3= 1-2\sigma$. {Denoting} $z_i = \sqrt{\lambda+\psi_i+\sqrt{(\lambda+\psi_i)^2-2\tilde{\epsilon}\chi\Sigma_i}}$, $\bar{z}_i = \sqrt{\lambda+\psi_i-\sqrt{(\lambda+\psi_i)^2-2\tilde{\epsilon}\chi\Sigma_i}}$ we may simplify the mean field equations \cite{supp}, which now reduce to

\begin{align}
	\frac{1}{2\ell^2\sqrt{2\chi}} 
	\left(\frac{2 \Sigma_1}{z_1+\bar{z}_1} +  \frac{\Sigma_3}{z_3+\bar{z}_3}\right)	-1  &=0 \\
	\frac{1}{2\sqrt{\tilde{\epsilon}}} 
	\left(\frac{2}{z_1+\bar{z}_1} +  \frac{1}{z_3+\bar{z}_3}\right)- 1 &=0 \\
	\frac{1}{2\sqrt{\tilde{\epsilon}}} \frac{1}{z_1+ \bar{z}_1}  -  \sigma &=0 \\
	\frac{1}{2\sqrt{\tilde{\epsilon}}} \frac{1}{z_3+ \bar{z}_3}  - (1-2 \sigma) &=0 \\
	-  \psi_1 + \tilde{u} c  \left(1-\sigma\right)  &=0 \\
	-  \psi_3 + \tilde{u} c  2\sigma  &=0 .
\end{align}

\noindent
After some algebra \cite{supp} 
we can combine these equations and 
get a {simple algebraic} equation for $\sigma$

\begin{align}\label{eq:mean field central2}
	\frac{1}{8 \sigma^2} - \frac{1}{8 (1-2\sigma)^2} - \tilde{u} c\tilde{\epsilon} (1-3\sigma) - \frac{\tilde{\epsilon}^2}{\ell^2} \left(\sqrt{\Sigma_1}-\sqrt{\Sigma_3}\right)\left[2\sigma \left(\Sigma_1-\Sigma_3\right)+\Sigma_3\right] &=0,
\end{align}

\noindent
which can be simplified a little bit more (seeing that $\ell$ is just a length scale, and that $c$ {appears only combined with} $u$) by redefining ${\epsilon} =\frac{\tilde{\epsilon}}{\ell}$, ${u} = \tilde{u} c \ell = \tilde{u} \frac{L}{2\pi \ell} $. We get

{\begin{align}\label{eq:mean field central}
	\frac{1}{8 \sigma^2} - \frac{1}{8 (1-2\sigma)^2} - {u} {\epsilon} (1-3\sigma) - {\epsilon}^2 \left(\sqrt{\Sigma_1}-\sqrt{\Sigma_3}\right)\left[2\sigma \left(\Sigma_1-\Sigma_3\right)+\Sigma_3\right] &=0.
\end{align}}

Eq. \eqref{eq:mean field central} is the central result of the paper, and all results stem from it. In order to compare different ellipsoids we solve this equation for oblate and prolate spheroids with the same area $A = 4\pi\ell^2$. {As mentioned above, $\ell$ is the "ellipsoid scale" and is roughly the geometric mean of the semi-major and semi-minor axes of the ellipsoid. While lacking any deep physical meaning, it does provide a convenient scale to relate to.}  In the following, when discussing the results, we will prefer to parametrize the form eigenvalues $\Sigma_1$ and $\Sigma_3$ via the flattening index, $\delta$ defined as \cite{newton1726philosophiae,bessel1825uber,bessel2009calculation,grossman2021packing},

\begin{align}
	\delta = \sqrt{\frac{\Sigma_1}{\Sigma_3}}-1.
\end{align}

\noindent
{In terms of the flattening we can express $\Sigma_1$ and $\Sigma_3$ as}

\begin{align}
	\Sigma_1&= \frac{1}{2} \left[1+ \frac{\left(1+\delta\right)^2\arccos\left(\frac{1}{1+\delta}\right)}{\sqrt{\delta \left(2+\delta\right)}}\right] ,
\end{align}

\noindent
and

\begin{align}
\Sigma_3&= \frac{1}{2 \left(1+\delta\right)^2} \left[1+ \frac{\left(1+\delta\right)^2\arccos\left(\frac{1}{1+\delta}\right)}{\sqrt{\delta \left(2+\delta\right)}}\right].
\end{align}

\section{Analysis}

As shall be seen, Eq. \eqref{eq:mean field central} may exhibit a number of solutions within the allowed  range, all of which are saddle-points of the free energy (as opposed to either minima or maxima), stemming from the mean-field approach adopted here. Nevertheless, the stability of the solutions within the mean field approach, can still be {determined} using catastrophe theory, in which we view {the} equation as stemming from a an effective potential \cite{saunders1980introduction,misbah2016complex}.  Multiplying Eq. \eqref{eq:mean field central} by $\sigma^2\left(1-2\sigma\right)^2$ results in a 5$^{th}$ order polynomial equation. Within catastrophe theory, this corresponds to the butterfly catastrophe model \cite{misbah2016complex,chirilus2015butterfly}. Typically, the butterfly model has $4$ free parameters ({namely the} coefficients of the {$1^{st}$} to {$4^{th}$} order monomials). In contrast, the system introduced here has only $3$ {parameters}. However, since the transformation between the canonical form of the butterfly catastrophe and our expression is very non-trivial, and sometimes diverging, we still recover a significant portion of the phenomenology of the butterfly model, that {cannot be captured by lower order catastrophes}. Yet, since physical solutions of $\sigma$ {from} Eq.\eqref{eq:mean field central} must lie within the range $0 \leq \sigma \leq \frac{1}{2}$, and the complexity of the transformation to a canonical form, little insight can be gained by using {the form suggested by catastrophe theory}, and we continue with the formulation given above. We thus write an effective potential $V_{\rm eff}(\sigma)$ such that Eq. \eqref{eq:mean field central} can be written as $-\frac{\pd V_{\rm eff}}{\pd \sigma}=0$, {namely it expresses the vanishing of the gradient of $V_{\rm eff}(\sigma)$. The} stability analysis is done via the derivative of Eq. \eqref{eq:mean field central}, while determination of metastability is done via comparing the actual free energy $W$, {from Eq. \eqref{eq: Free energy MF}}, of the  {different} solutions.

\begin{figure}
\centering
\includegraphics[width=\textwidth]{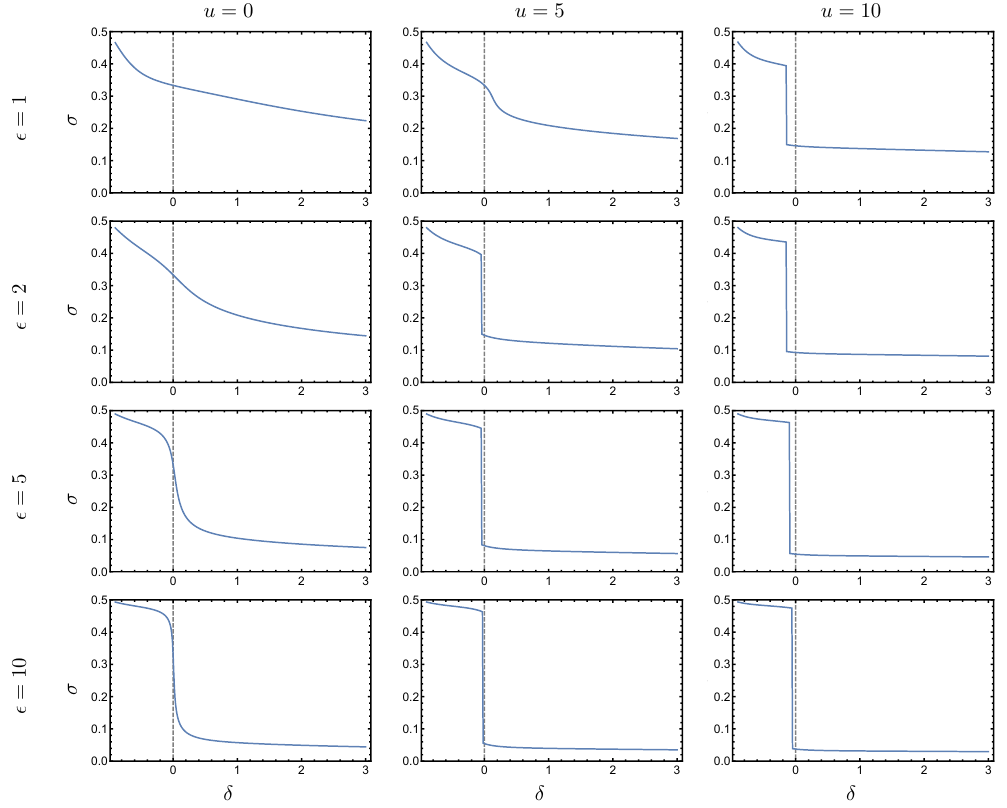}
\caption{$\sigma$ vs $\delta$, for different values of the persistence length $\epsilon$ (rows) and the self-avoidance strength $u$ (columns). 
The vertical gray dashed line marks the case of a sphere ($\delta =0$). 
Longer persistence lengths drive the system into more ordered states
along the large dimension of the ellipsoid. Stronger self-avoidance, however, both adds a metastable state (not shown in this figure, but discussed in \cite{supp}), 
and drive the lowest energy state to order along the poles. This is a purely entropic effect, caused by the existence of more ordered states along this direction. This happens even for oblate spheroids ($\delta<0$). Note that the $\delta = 0$ case, corresponding to a sphere, should be viewed as the limit $\delta \rightarrow 0$, since the treatment here assumes from the beginning a broken symmetry (namely an axial symmetry and not a spherical one).
\label{fig: sig_vs_delta}}
\end{figure}

\begin{figure}
\centering
\includegraphics[width=\textwidth]{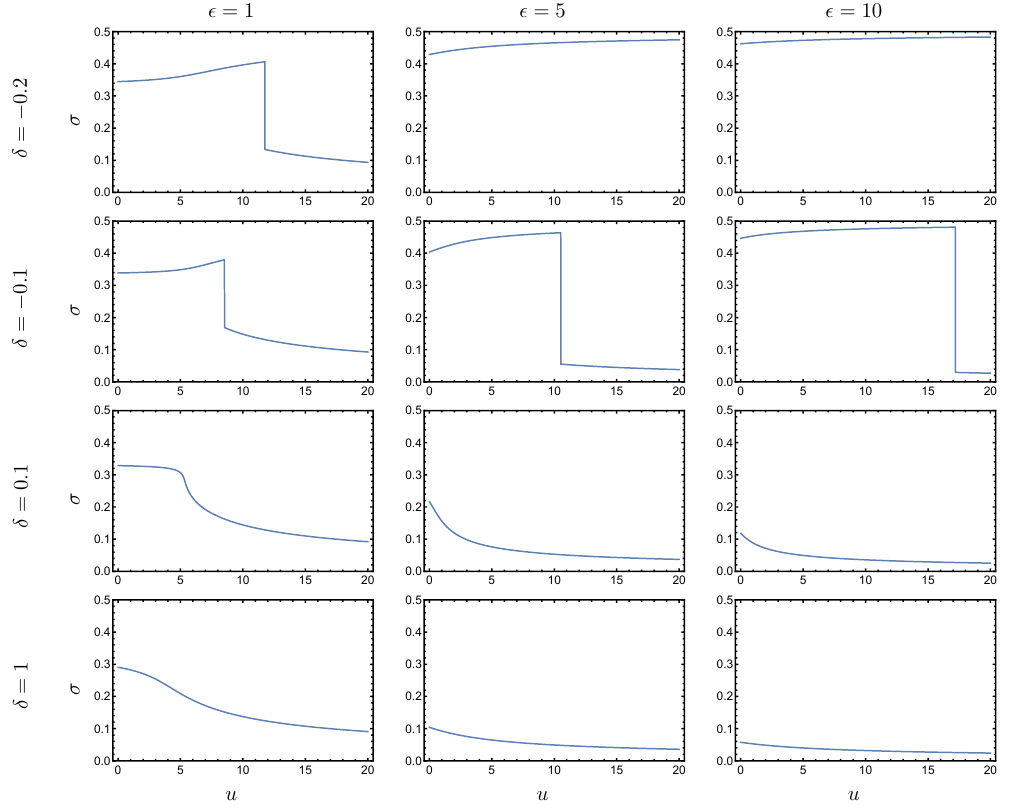}
\caption{$\sigma$ vs $u$, for different values of the flattening  $\delta$ (rows), and persistence lengths $\epsilon$ (columns). In strong self-avoidance, the system is driven into a polar oriented configuration even for oblate spheroids. In this case the transition is an abrupt one (namely a first order transition) due to entropy.
\label{fig: sig_vs_u}}
\end{figure}

\begin{figure}
\centering
\includegraphics[width=\textwidth]{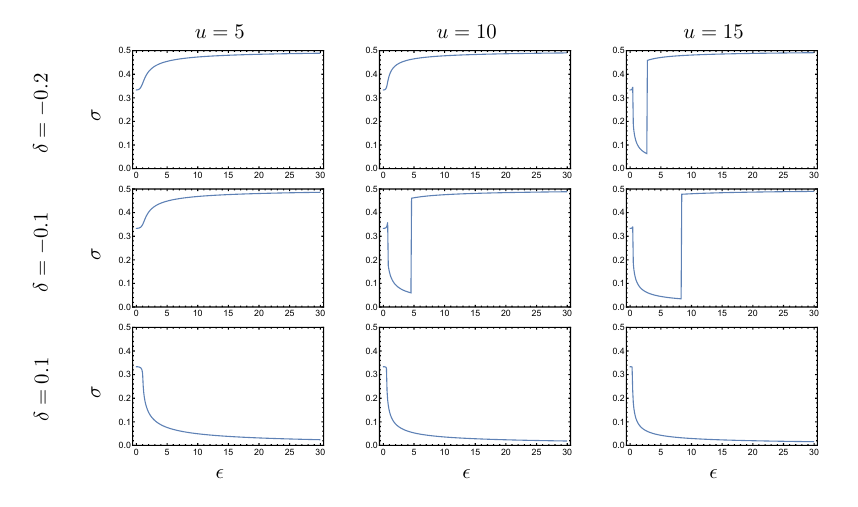}
\caption{$\sigma$ vs $\epsilon$, for different values of the flattening $\delta$ (rows) and self-avoidance $u$ (columns). The longer the persistence length, the energy becomes more important (rather than entropy), and the filaments align with the principal axis of the ellipsoid.  More specifically, for $\delta<0$ the transition (from entropy dominated to energy dominated phases) is of first order.
\label{fig: sig_vs_epsilon}}
\end{figure}

Figures \ref{fig: sig_vs_delta} - \ref{fig: sig_vs_epsilon} show the different {least energy} states of Eq. \eqref{eq:mean field central} while varying either $\delta$, $u$, or $\epsilon$ for given values of the other two parameters. Most notably, we can identify a very different behaviour depending on the flattening, $\delta$. When $\delta>0$ (prolate ellipsoid) a gradual (though maybe fast) transition into a polar order is seen as either $\delta$, $u$, or $\epsilon$ get larger. This is a very natural result as we would expect stiffer filaments to prefer less bent configurations. These can be achieved by ordering along the longer, lower average curvature, direction. A similar argument is valid for {large} flattening, in which most of the ellipsoid has a very small curvature along {its long axis}, except for a small region near the poles where the curvature is very large. In contrast, the effect of self-avoidance is more subtle. Naively, one would expect that when self-avoidance is very strong certain non-intersecting configurations will become more stable even if they are not energetically preferential. Such possible configurations include both a polar order (described by $\sigma \sim 0$), {in which the filament mostly points toward the poles}, but also an azimuthal order (described by $\sigma \sim 0.5$), {in which the filament points mainly along the orthogonal direction to the ellipsoid axis of symmetry}. As such, except for driving the ground state faster into a more ordered state (as can be seen, e.g in Figure \ref{fig: sig_vs_u}), it also gives rise to a meta-stable state with $\sigma \sim 0.5$ \cite{supp}. 
The range of parameter in which this state exists depends both on $\delta$ and $\epsilon$ since, for example, filaments that are very stiff (or very soft),  will snap (or easily "slide") out of this metastable arrangement.

An oblate spheroid, describe by $\delta <0$, exhibits a very different behaviour. When {there is no self-avoidance} $u=0$, the ground state continuously becomes more ordered along the azimuthal direction ($\sigma = 0.5$) both as $\epsilon$ gets larger (stiffer filaments) and as $\delta$ gets closer to $-1$ ({namely, a} flatter ellipsoid). However, when self-avoidance is strong relative to the persistence length{,} an abrupt, discontinuous transition occurs. Suddenly, the order parameter jumps from $\sigma \sim 0.5$  to $\sigma \sim 0$. This transition occurs since for strong self-avoidance (or high densities), entropy plays an important role, similar to the role of entropy in hard-sphere packing. In such a case, filaments will strongly align, and due to the broken symmetry of the problem there are many more available non-intersecting configurations aligning {along the polar direction (polar order)} rather than the azimuthal where there is essentially a single configuration with low energy. {This implies that polar packings enjoy larger entropy than azimuthal packings.}
An important issue is that the $\delta =0$ case, {which corresponds to a sphere}, should be understood here as the limit $\delta \rightarrow 0$. This is because {our discussion assumes a} finite value of $\delta$, or in simple words assumes that the symmetry of the container is broken. Setting $\delta =0$ in Eq. \eqref{eq:mean field central} one can easily see that $\sigma = 0.33$ is the only solution in this case, which under global averaging does not differentiate between ordered or disordered states (due to {the} high symmetry of {the sphere}). 

As expected, for a given $\epsilon$, the aforementioned jump of the ground state occurs at lower values of $\delta$ as $u$ gets stronger (see Fig. \ref{fig: sig_vs_delta}). Similarly, at {a} given  $\delta${,} the critical $u$ at which the transition occurs, grows with $\epsilon$ (see Fig. \ref{fig: sig_vs_u}). Finally, as in any first order transition, metastable states appear around it, and we leave a more comprehensive discussion of them for \cite{supp}.


\begin{figure}[h!] 
\centering
\includegraphics[width=0.6\textwidth]{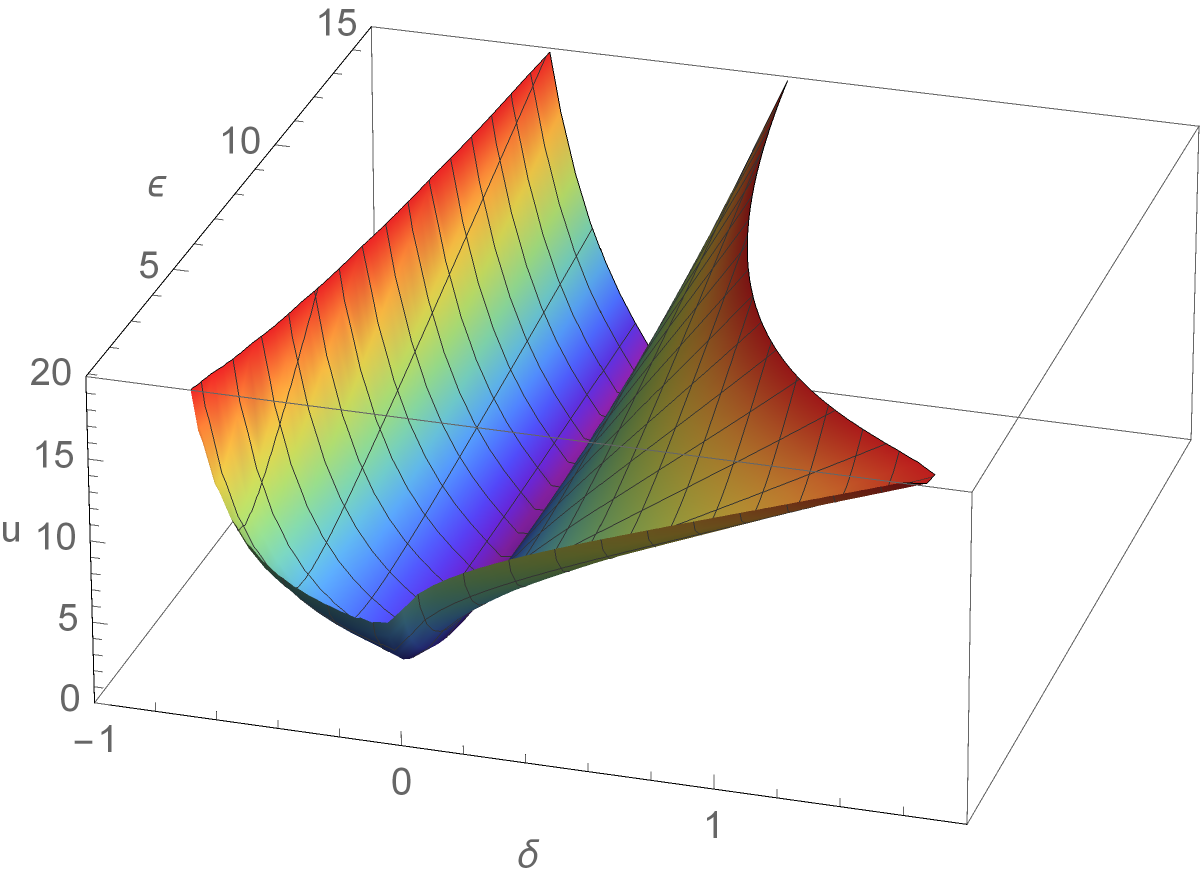}
\caption{In this figure we present the critical surface $u^*(\delta,\epsilon)$. The volume above this surface describes the region in parameter space, where there are $3$ physical solutions, namely one stable, one metastable, and one unstable, while underneath it there is only one (stable) solution. Note that this surface is infinite, in the sense that for every $\delta$ and $\epsilon$ there exists a $u^*$ - above which there are $3$ solutions.
\label{fig:crit_surf}}
\end{figure}

Formally, Eq. \eqref{eq:mean field central} may have up  to $5$ solutions, however only up to $3$ are physical {solutions} (one is the ground state solution, one metastable and one unstable), satisfying the constraint $0 \leq \sigma \leq \frac{1}{2}$. Fig. \ref{fig:crit_surf} depicts the critical surface $u^*(\epsilon,\delta)$. For any $u<u^*$ there is only one (stable) solution, while for $u>u^*$ there are three solutions, where two are locally stable (i.e. one globally stable and one metastable), while one is unstable.

In order to provide further insight, we consider a completely random filament over the ellipsoid, to eliminate effects that relate only to the geometry of the surface rather than the elastic and entropic response of the filament to that geometry. A direct calculation of this dependence \cite{supp} 
yields 

\begin{align}
\sigma_{\rm rand} = \frac{\sqrt{\delta\left(2+\delta\right)}\left[\delta\left(2+\delta\right) - 1\right] + \left(1 + \delta\right)^4 \arccos\left(\frac{1}{1+\delta}\right)}{4 \delta \left(2 + \delta \right)\left[\sqrt{\delta \left(2 + \delta\right)} + \left(1 + \delta\right)^2 \arccos\left( \frac{1}{1+\delta} \right) \right]}.
\end{align}

\begin{figure}[h] 
\centering
\includegraphics[width=0.5\textwidth]{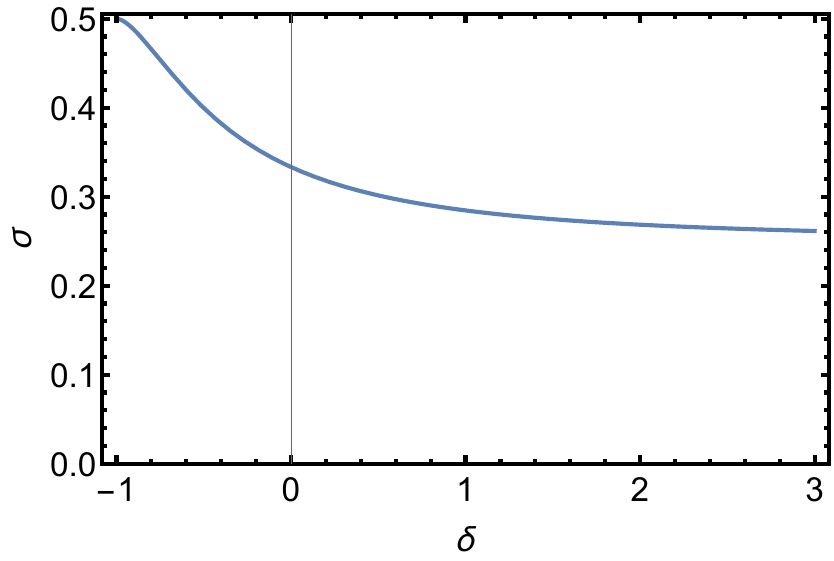}
\caption{The global order parameter $\sigma_{\rm rand}$ of a randomly oriented director field on the surface of an ellipsoid, as a function of {its flattening} $\delta$. $\sigma_{\rm rand} \left(\delta \rightarrow -1\right) = \frac{1}{2}$ as expected from a random director field on a disc. $\sigma_{\rm rand} \left(\delta \rightarrow \infty\right) = \frac{1}{4}$ as expected from a random director field on a cylinder.
\label{fig:random_ord}}
\end{figure}

\noindent
$\sigma_{\rm rand}$ is plotted in Fig. \ref{fig:random_ord}. As can be clearly seen, at the $\delta \rightarrow -1$ limit $\sigma_{\rm rand} \rightarrow \frac{1}{2}$. In this limit the oblate spheroid turns into a disc on the X-Y plane, which is in accordance with the result. At the $\delta \rightarrow \infty$ limit the spheroid is actually a cylinder, in which case the expected value for $\sigma_{\rm rand}$,  is $\sigma_{\rm rand}(\infty)= \frac{1}{4}$ {, and when $\delta = 0$, (corresponding to a sphere) $\sigma_{\rm rand} = \frac{1}{3}$, as result of the high degree of symmetry}. From Fig. \ref{fig:random_ord} it is clear that the effects seen in Figures \ref{fig: sig_vs_delta} - \ref{fig: sig_vs_epsilon} are not simply geometric, and are related to the energy and entropy of the configuration, since the value of $\sigma$ is driven below $\sigma_{\rm rand}$.

The transitions described here are expected to affect the global physical response and behavior of the system, e.g. the filament imposes stresses on the surface and may deform it. Therefore, we calculated the pressure $P$ exerted by the filament on the walls of the ellipsoid (associated with a change in the area), the shear force $T$ (associated with a change in the flattening), and the injection force $F$ needed to further inject the filament into the ellipsoid (associated with a change in the length of the filament). For solutions of Eq. \eqref{eq:mean field central}, one can express the energy $W$ as

\begin{align}\label{eq: energy_L_A}
	W = &\frac{\sqrt{2\pi} L}{16\sqrt{A} \epsilon} \Bigg\{\frac{1}{\left(1-2\sigma\right)^2} + \frac{1}{\sigma^2} - 8 u \epsilon \left(1-2\sigma\right)\left(1-9\sigma\right)   \\ \nonumber& +8 \epsilon^2\left[6\sigma\left(\Sigma_1-\Sigma_3\right)+3\Sigma_3 - \left(\sqrt{\Sigma_1}+\sqrt{\Sigma_3}\right)\right]\left[2\sigma\left(\Sigma_1-\Sigma_3\right)+\Sigma_3\right]\Bigg\} \\ \nonumber
	= & \frac{\sqrt{2\pi} L}{16\sqrt{A} \epsilon} \mathcal{W},
\end{align}

\noindent
where we defined $\mathcal{W}$ as the energy density per unit length of the filament.

\begin{figure}[h] 
\centering
\includegraphics[width=\textwidth]{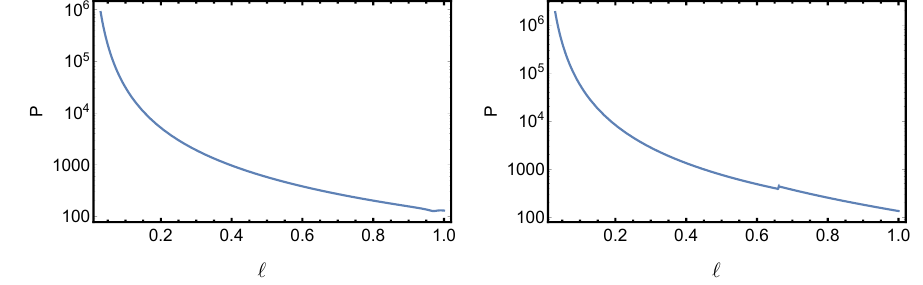}
\caption{{The} pressure {of the stable solution}, exerted on the ellipsoid by the filament, as a function of the ellipsoid scale $\ell =\sqrt{\frac{A}{2\pi}}$,  provided by Eq. \eqref{eq: pressure}. Left: $\delta =0.1$,  right: $\delta= -0.1$. In both graphs $L= 10$, $\tilde{u}=1$  and $\tilde{\epsilon}=1$. The small jump in the pressure in the right figure corresponds to the jump in the stable solutions  from polar-like ($\sigma \sim 0$) to equatorial-like ($\sigma \sim 0.5$) configurations as $\ell$ gets closer to $1$. {This is} similar to the transitions that can be seen in  Figs. \ref{fig: sig_vs_u} and \ref{fig: sig_vs_epsilon}, as $u$ and $\epsilon$ decrease.\label{fig:pressures} }
\end{figure}

The pressure $P$, exerted by the filament on the walls of the {ellipsoid}, is given by

\begin{align}\label{eq: pressure} 
	P =& \frac{d W}{d A} =\frac{\sqrt{2\pi} L}{16\sqrt{A} \epsilon} \frac{d \mathcal{W}}{d A} = -\frac{\sqrt{2\pi} L}{32\sqrt{A^3} \epsilon}\left( \frac{ d \mathcal{W}}{d u} u + \frac{ d \mathcal{W}}{d \epsilon}\epsilon\right) ,
\end{align}

\noindent
where $\frac{d \mathcal{W}}{d u} = \frac{\pd \mathcal{W}}{\pd u} + \frac{\pd \mathcal{W}}{\pd \sigma} \frac{d \sigma}{d u}$ and similarly for $\epsilon$.
In Fig. \ref{fig:pressures} we plot the pressure for a prolate ($\delta = 0.1$, left) and oblate ($\delta = -0.1$, right) spheroid, as a function of the ellipsoid scale $\ell = \sqrt{\frac{A}{2\pi}}$, for $\tilde{u}=1$, $\tilde{\epsilon}=1$, and $L=10$, $\ell= \frac{1}{2}$. Due to the scaling of the problem, as $\ell$ gets larger we simultaneously increase $u$ and decrease $\epsilon$. For this set of parameters, $\ell=1$ corresponds to an ellipsoid whose length-scale is similar to the persistence length, and thus the mean-field approximation is expected to break. {We therefore do not plot this figure beyond $\ell=1$}. Similar plots for other values of the parameters result in similar looking graphs, though obviously the {specific} numbers change. For the case of $\delta = -0.1$, a small jump in pressure can be seen around $\ell \sim 0.7$. This jump corresponds to the snap-through of the configuration from an azimuthal order (described by $\sigma \sim 0.5$) for small $\ell$'s to polar order (described by $\sigma \sim 0$) for large $\ell$'s, where entropy governs the behaviour of the system.

\newcommand{\wid}{0.47\textwidth}
\newcommand{\swid}{0.9\textwidth}

The normalized elongation shear stress $T$ {is given by}

\begin{align}\label{eq: shear stress}
T =& -\frac{d W}{d \delta} =-\frac{\sqrt{2\pi} L}{16\sqrt{A} \epsilon} \frac{d \mathcal{W}}{d \delta}.
\end{align}

\begin{figure}[h]
\centering
\includegraphics[width=\textwidth]{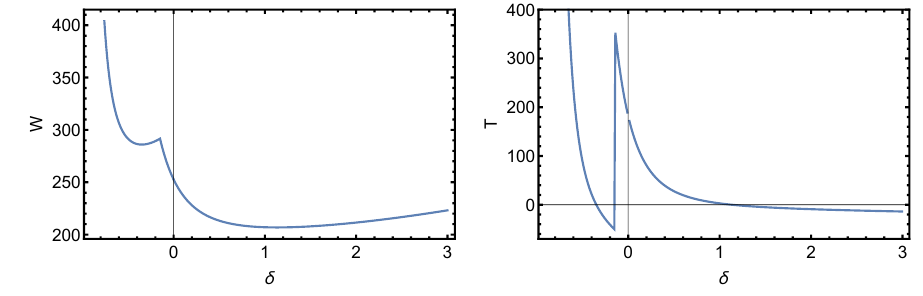}
\caption{Left: {The free energy $W$} of the ground state as function of $\delta$ {given by Eq. \eqref{eq: energy_L_A}}, and Right: the corresponding elongation shear stress $T$ exerted on the ellipsoid for deformations with same area {given by Eq. \eqref{eq: shear stress}}. In both graphs $\tilde{u}=\tilde{\epsilon}=1$, $L=10$, $\ell=\frac{1}{2}$ which correspond to $u=6.36$ and $\epsilon=2$. Note a cusp in the energy (and jump in {the} shear stress) at the critical $\delta_c$ where the solution jumps from an equatorial-like ($\sigma \sim 0.5$) to polar-like ($\sigma \sim 0$) configurations. Additionally, it {is} clear that in both cases, namely $\delta<0$ and $\delta>0$, there are some stable solutions, {in which the energy is at a local minimum, if the ellipsoid shape is allowed to change.}
\label{fig:shear} }
\end{figure}

\noindent
The elongation shear stress $T$ is plotted in Fig. \ref{fig:shear} (right) as a function of $\delta$ alongside the free energy $W$ (left), for the case $\tilde{u}=1$, $\tilde{\epsilon}=1$, and $L=10$. It can be clearly seen that there are two values of $\delta$ for which the configuration is stable to shear. One is at $\delta<0$, which corresponds to an azimuthal order, and one for $\delta>0$ corresponding to polar order. These two shear-stable solutions, are separated by a cusp at some {negative} $\delta_c$ corresponding to the {cusp in the free energy}. These results {suggest} that for $\delta>\delta_c$, if we allow the ellipsoid to change {its} flattening (and only this property), it will change its shape to the locally minimizing one (assuming the filament ordering relaxes fast enough). The exact position of the local minima, as well as the cusp depend on the exact values of the problem, but {we expect} that the qualitative result remains the same. Interestingly, for vanishing $\tilde{u}$ (namely in the absence of self-avoidance) the cusp turns into a local maximum.

Finally, the injection force $F$ is given by

\begin{align}\label{eq: force}
	F =& \frac{d W}{d L} =\frac{\sqrt{2\pi} }{16\sqrt{A} \epsilon} \mathcal{W} +\frac{\sqrt{2\pi} L }{16\sqrt{A} \epsilon}\frac{d \mathcal{W}}{d L} =
	\frac{\sqrt{2\pi} }{16\sqrt{A} \epsilon} \mathcal{W} +\frac{\sqrt{2\pi} u }{16\sqrt{A} \epsilon}\frac{d \mathcal{W}}{d u} 	.
\end{align}

\begin{figure}[h]
\centering
\includegraphics[width=\textwidth]{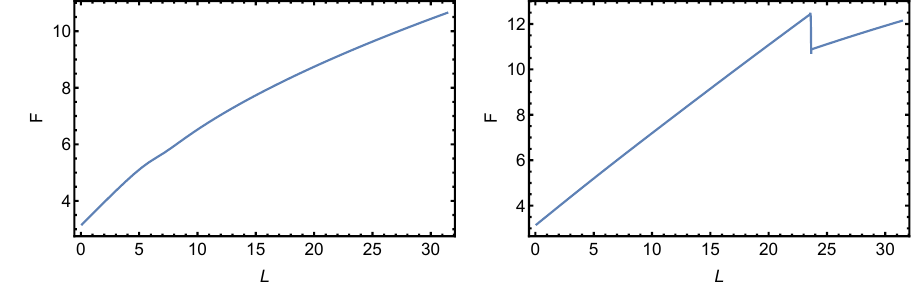}
\caption{{The injection} force $F$ needed to push further the filament into the ellipsoid, {given by Eq. \eqref{eq: force}}, as {a} function of {the} total length of the filament, $L$,  already inside the container. Left: $\delta =0.1$ (prolate),  right: $\delta= -0.1$ (oblate). In both graphs $\ell= \frac{1}{2}$ (and hence $A =\frac{\pi}{4}$), $\tilde{u}=1$  and $\tilde{\epsilon}=1$ (thus, $\epsilon=2$). The jump in the force in the right figure corresponds to the jump in stable solutions from  equatorial-like ($\sigma \sim 0.5)$ to polar-like ($\sigma \sim 0$) as $L$ increases. {This is} similar to the transitions that can be seen in Figs. \ref{fig: sig_vs_u} as $u$ increases.  
\label{fig:forces} }
\end{figure}

\noindent
The injection force $F$ is plotted in Fig. \ref{fig:forces} for prolate ($\delta = 0.1$, left) and oblate ($\delta = -0.1$, right) {ellipsoids}, as a function of the filament length $L$, for $\ell = \sqrt{\frac{A}{2\pi}} = \frac{1}{2}$, $\tilde{u}=1$, and $\tilde{\epsilon}=1$. As the length $L$ gets longer $u$ is increased, and we therefore see that in the case of an oblate {ellipsoid ($\delta<0$)}, {and for} a long enough filament, a jump is observed in the configuration and respectively in the injection force. {This comes from the fact that} the filament changes its configuration into a polar one at high densities ({an} entropy driven snap-through) {as discussed also in Fig. \ref{fig: sig_vs_u}}.  Other values of $\ell$, $\tilde{u}$ and $\tilde{\epsilon}$, and $\delta$  result in similar graphs, though naturally the specific numbers change. In the case of vanishing $\tilde{u}$ no jump is seen as the system is energetically driven, and not affected by entropic considerations.

\section{Discussion}

{All the calculations done in this paper are limited to the mean-field approximation and further limited by the global average made to simplify the calculation. One might fear that these simplifications may lead to loss of significant (i.e qualitative) effects related to the shape of the container. However, this is not the case, as clearly seen from the results which strongly depend on the flattening index $\delta$. This dependence remains due to two reasons. First, the mean-field approximation is done with respect to the filament, not the shape of the container. This means that any forcing that confines the filament to the surface of the container (and imposed using a Lagrange multiplier) is true only on average. When the persistence length is small, the filament, under this approximation, may not adhere completely to the surface. In any case, when making this approximation (leading to Eq. \eqref{eq: mean field calc}), the order parameter is still given by a local average. The second reason that enables us to keep track of the geometry of the container is the fact that the global averaging is done in the embedding 3D space. In fact, the shape affects this average in a cardinal way, as can be seen in Fig. \ref{fig:random_ord}, in which the locally random field is the result of an average over the shape of the ellipsoid. This effect is purely geometrical. Finally, with regard to mean-field approximations in general, while the actual quantitative results may differ from those reported here, we expect that a faithful qualitative description is indeed provided. This is a robust feature of mean-field approaches in critical phenomena. \blu{In fact, these approximations are expected to be exact in the $\delta\rightarrow0$ limit, in the sense that on a sphere, there is no approximation done in the calculation of the averages.}}

The results presented in this paper show the statistics of self-avoiding semi-flexible filaments packed on the surface of ellipsoids, both oblate ($\delta<0$) and prolate ($\delta>0$).  We find that the geometry plays a significant role in the orientation of the filaments, and that without self-avoidance, filaments tend to naturally order along the large dimension of the ellipsoid. This result is expected intuitively and supports the results {derived} in Ref. \cite{grossman2021packing}. {Considering} self-avoidance adds both energetic and entropic {contributions} that give rise to surprising metastable solutions, and a rich behaviour of filament ordering. Most notably, an entropy induced snap-through transition is expected when varying the ellipsoid's aspect ratio. Such entropic effects were described in the packing of DNA inside bacteria \cite{jun2006entropy}. {In essence, due to self-avoidance, the filaments cannot roll around the equator indefinitely, and at a certain critical (negative) flattening $\delta_c<0$, they pick up a polar configuration.}

Despite describing the order of the filaments using an order parameter $\sigma$, the fact {that} it is a global, rather than a local measure, means that this work lacks a detailed description of the {filament configurations in the ordered phase}. Thus, it is hard to compare it to results such as those reported in 
\cite{mackintosh1991orientational,lubensky1992orientational,evans1995phase,vitelli2006nematic,gerlach2011longest,cerda2005excluded}{, studying the actual configuration of nematic orderings}.  Nevertheless, an important difference is that the filament length in the system under study here is very long, which stand in contrast to the expected behavior of short segments, the latter should behave similar to nematics.  Finite-length effect were not taken into account here and are likely, at least in low-to-intermediate densities, to play a role, especially in filaments whose length is comparable to the ellipsoids' size.

{The results also support effects seen in other works regarding the configurations of flexible filaments in non-trivial containers, such as those in \cite{marenduzzo2008computer,marenduzzo2009statistics,marenduzzo2010biopolymer,ali2006polymer,petrov2007conformation,petrov2008packaging,petrov2009characterization,vetter2013finite,vetter2014morphogenesis,vetter2015growth, pineirua2013spooling,purohit2003mechanics,angelescu2008viruses,fritsche2011confinement}, and serve to both strengthen the importance of the container geometry, and expand those works by adding analytical insight to the complex phase-space that results from considering both geometry and self-avoidance.}

\begin{figure}
\centering
\includegraphics[width=0.5\textwidth]{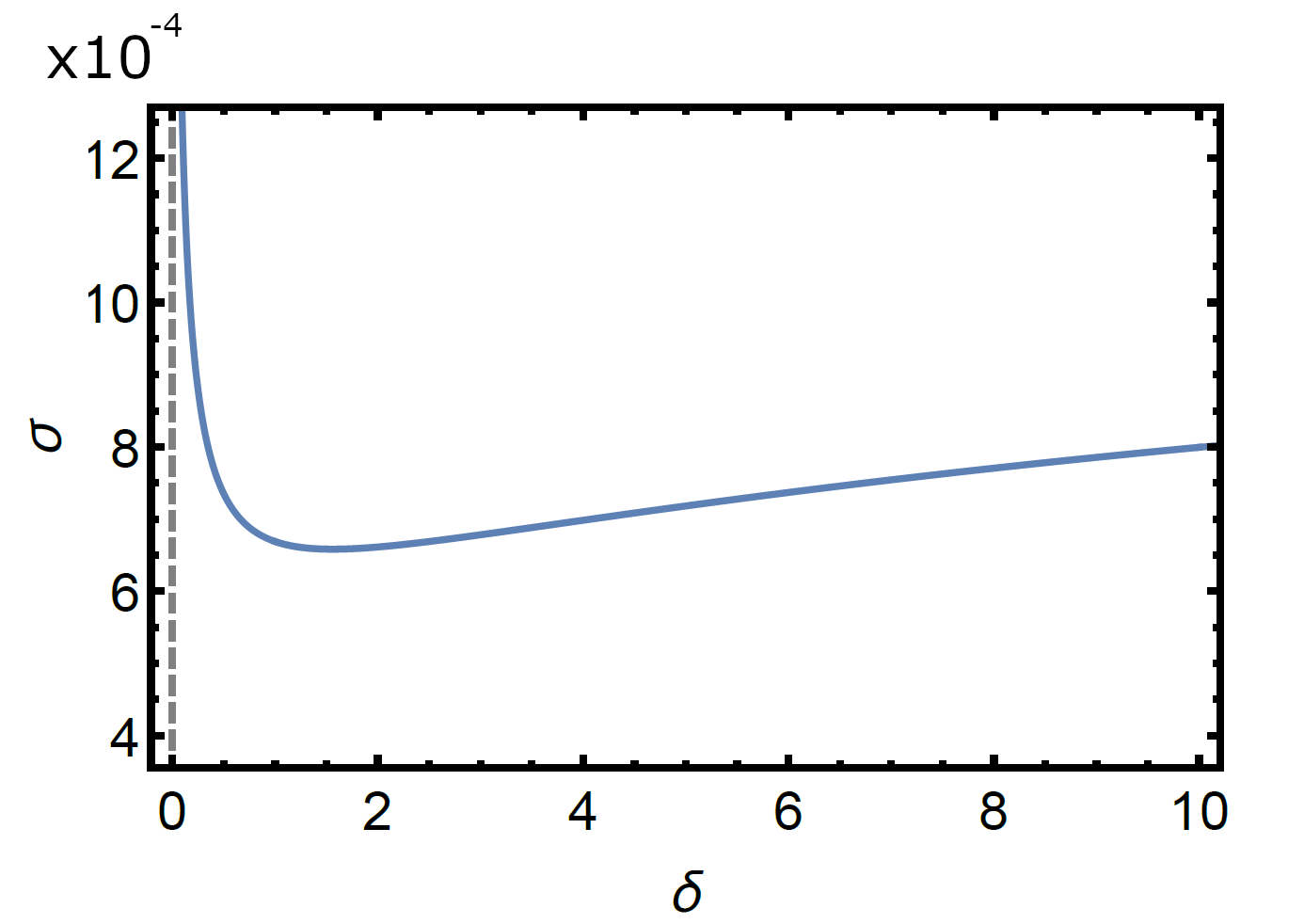}
\caption{$\sigma$ as function of $\delta$ for $u=0$ and $\epsilon=10,000$, approaching the limit of pure cold, non-self-avoiding filaments. The fact that $\sigma$ has a local minimum at small $\delta$'s suggests that a transition similar to the one described in Ref. \cite{grossman2021packing} occurs.\label{fig: low_temp} }
\end{figure}

It is worth noting that at high persistence lengths, a transition reminiscent of the one suggested in Ref. \cite{grossman2021packing} can be seen - see Fig. \ref{fig: low_temp} for an example. As a function of $\delta$, we find a minimal value of $\sigma$, beyond which, as $\delta$ grows, $\sigma$ grows again. In Ref. \cite{grossman2021packing}, the configurations of phantom (i.e. non self-avoiding) filaments on an ellipsoid were studied from a purely mechanical (rather than statistical) point of view. It was found that in prolate ellipsoids ($\delta>0$), the filaments prefer, a configuration that passes through the poles, up until a critical $\delta^*$ is reached. Beyond this critical value $\delta^*$, the configurations break the perfect alignment with the ellipsoid's major axis, and instead choose a configuration that bypasses the poles. This happens since for large values of $\delta$, the curvature at the poles becomes large. The main difference between the current observation, and the transition originally suggested, is that this observation is a smooth and not an abrupt transition.  This result suggests that the high curvature at the poles indeed plays some role in any actual configuration that might be observed. Namely, these filament configurations try to avoid the poles when $\delta$ is large.

To conclude, this paper suggests that geometry plays a significant, and often {given little attention}, role in the alignment of filaments {on} anisotropic containers. {In that respect,} the results shown here stand on their own, but are also complimentary to those presented in Ref. \cite{grossman2021packing}, {by} treating {a} more complicated  {situation, namely including self-avoidance, yet at the price of not providing access to} possible configurations. {Another important conclusion is that self-avoidance also plays a significant role by reducing the phase space of possible configurations, such that under certain conditions the system is forced to take a radically different configuration, compared to those of a phantom chain. This is the mechanism underlying the snap-through (entropy driven) transition reported above}.

Finally, this work does not consider asymmetric spheroids or other shapes of containers \cite{marenduzzo2008computer,marenduzzo2009statistics,marenduzzo2010biopolymer,ali2006polymer,petrov2007conformation,petrov2008packaging,petrov2009characterization,vetter2013finite,vetter2014morphogenesis,vetter2015growth, pineirua2013spooling,purohit2003mechanics,angelescu2008viruses,fritsche2011confinement}, charged filaments
\cite{netz2003neutral,borukhov1997steric,abrashkin2007dipolar,slosar2006connected,van2005electrostatics,ghosh2002phase,carri1999attractive,fathizadeh2013confinement,angelescu2008viruses}, filaments that exhibit torsion \cite{stoop2011packing,shaebani2017compaction} or plasticity \cite{shaebani2017compaction},
dynamics \cite{purohit2003force,purohit2005forces,kindt2001dna,smith2001bacteriophage,fathizadeh2013confinement}, and is limited to a mean-field approximation. As such, there is much more work to be done. The inclusion of friction, as done in Refs. \cite{stoop2011packing,shaebani2017compaction} for example, will likely produce hysteresis depending on the exact way in which filaments are inserted into the container. Actually, the case of filaments that extend into the volume is very interesting in its own right \cite{stoop2011packing,shaebani2017compaction, morrison2009semiflexible,liang2019orientationally} and it is worth exploring using the methods used here.
Attempting to advance beyond the mean-field approximation is likely to improve the accuracy of the results presented here for the pressure and injection force and furthermore, to yield information regarding the configuration of the filaments. A promising approach is the self-consistent expansion \cite{SCE2008}, which is known to improve over variational and mean-field approximations.



%

\clearpage
\appendix
\section{Detailed Calculations}\label{app: detailed derivation}

In this appendix we provide further details that lead to the simplification of the mean-field equations (18)-(21).
Consider integrals of the form 

\begin{align}
	&\int\limits_0^\infty \frac{\dt q}{\tilde{\epsilon} q^4/2 + (\lambda+\psi) q^2 +\chi} = \frac{1}{\tilde{\epsilon}} \int\limits_{-\infty}^{\infty} \frac{\dt q}{q^4 + 2(\lambda'+\psi') q^2 + \chi'} = \\ 
	\nonumber &= \frac{1}{\tilde{\epsilon}}\int\limits_{-\infty}^{\infty}  \frac{\dt q}{\left(q-i \sqrt{\beta + \sqrt{\beta^2 -\gamma}}\right)\left(q+i \sqrt{\beta + \sqrt{\beta^2 -\gamma}}\right)\left(q-i \sqrt{\beta -\sqrt{\beta^2 -\gamma}}\right)\left(q+i \sqrt{\beta - \sqrt{\beta^2 -\gamma}}\right)}
\end{align}
where we defined $\lambda' = \lambda/\tilde{\epsilon}$, $\psi'=\psi/\tilde{\epsilon}$, $\chi'= 2 \chi/\tilde{\epsilon}$, and $\beta =\lambda'+\psi'$, $\gamma=\chi'$.

\noindent
Using the residue theorem we can calculate the following integrals

\begin{align}
	\int\limits_0^\infty \frac{\dt q}{\tilde{\epsilon} q^4/2 + (\lambda+\psi) q^2 +\chi} &=  \frac{\pi}{\sqrt{2\chi}\left(\sqrt{\lambda+\psi+\sqrt{(\lambda+\psi)^2-2\tilde{\epsilon}\chi}}+\sqrt{\lambda+\psi-\sqrt{(\lambda+\psi)^2-2\tilde{\epsilon}\chi}}\right)} , \\
	\int\limits_0^\infty \frac{ q^2 \dt q}{\tilde{\epsilon} q^4/2 + (\lambda+\psi) q^2 +\chi} &=  \frac{\pi}{\sqrt{\tilde{\epsilon}}\left(\sqrt{\lambda+\psi+\sqrt{(\lambda+\psi)^2-2\tilde{\epsilon}\chi}}+\sqrt{\lambda+\psi-\sqrt{(\lambda+\psi)^2-2\tilde{\epsilon}\chi}}\right)} .
\end{align}

\noindent
Thus, we are finally left with the following mean-field equations
\begin{align}
	\frac{1}{2} 
	\sum\limits_{i}\left[\frac{\Sigma_i}{\sqrt{2\chi}\left(\sqrt{\lambda+\psi_i+\sqrt{(\lambda+\psi_i)^2-2\tilde{\epsilon}\chi\Sigma_i}}+\sqrt{\lambda+\psi_i-\sqrt{(\lambda+\psi_i)^2-2\tilde{\epsilon}\chi\Sigma_i}}\right)}\right]	-\ell^2  &=0 \\
	\frac{1}{2}\sum\limits_{i}\left[\frac{1}{\sqrt{\tilde{\epsilon}}\left(\sqrt{\lambda+\psi_i+\sqrt{(\lambda+\psi_i)^2-2\tilde{\epsilon}\chi\Sigma_i}}+\sqrt{\lambda+\psi_i-\sqrt{(\lambda+\psi_i)^2-2\tilde{\epsilon}\chi\Sigma_i}}\right)}\right]- 1 &=0 \\
	\frac{L}{2} \left[\frac{1}{\sqrt{\tilde{\epsilon}}\left(\sqrt{\lambda+\psi_i+\sqrt{(\lambda+\psi_i)^2-2\tilde{\epsilon}\chi\Sigma_i}}+\sqrt{\lambda+\psi_i-\sqrt{(\lambda+\psi_i)^2-2\tilde{\epsilon}\chi\Sigma_i}}\right)}\right] - A \sigma_{i} &=0 \\
	-  \psi_{i} + \tilde{u}  \left((\sigma_1+\sigma_2+\sigma_3)-\sigma_{i}\right)  &=0 .
\end{align}

By defining $z_i = \sqrt{\lambda+\psi_i + \sqrt{(\lambda+\psi_i)^2-2\tilde{\epsilon}\chi\Sigma_{i}}}$ and $\bar{z}_i = \sqrt{\lambda+\psi_i - \sqrt{(\lambda+\psi_i)^2-2\tilde{\epsilon}\chi\Sigma_{i}}}$, and using the fact that $\sigma_1+\sigma_2+\sigma_3=1$ and $\sigma_1 = \sigma_2$ 
we get the simpler set of equations

\begin{align}
	\frac{1}{2\sqrt{2\chi}} 
	\left(\frac{2 \Sigma_1}{z_1+\bar{z}_1} +  \frac{\Sigma_3}{z_3+\bar{z}_3}\right)	-\ell^2  &=0 \\
	\frac{1}{2\sqrt{\tilde{\epsilon}}} 
	\left(\frac{2}{z_1+\bar{z}_1} +  \frac{1}{z_3+\bar{z}_3}\right)- 1 &=0 \\
	\frac{L}{2\sqrt{\tilde{\epsilon}}} \frac{1}{z_i+ \bar{z}_i}  - A c \sigma_{i} &=0 \\
	-  \psi_{i} +\tilde{u} c  \left(1-\sigma_{i}\right)  &=0 .
\end{align}

\noindent
This can further simplified into dimensionless quantities  by dividing Eq. (A8) by $\ell^2$ and Eq. (A10) by $Ac$, 
while keeping in mind that $c=L/A$

\begin{align}
	\frac{1}{2\ell^2\sqrt{2\chi}} 
	\left(\frac{2 \Sigma_1}{z_1+\bar{z}_1} +  \frac{\Sigma_3}{z_3+\bar{z}_3}\right)	-1  &=0 \\
	\frac{1}{2\sqrt{\tilde{\epsilon}}} 
	\left(\frac{2}{z_1+\bar{z}_1} +  \frac{1}{z_3+\bar{z}_3}\right)- 1 &=0 \\
	\frac{1}{2\sqrt{\tilde{\epsilon}}} \frac{1}{z_i+ \bar{z}_i}  -  \sigma_{i} &=0 \\
	-  \psi_{i} + \tilde{u} c  \left(1-\sigma_{i}\right)  &=0 
\end{align}

\noindent 
Inserting equations (A14) in equation (A12) we get

\begin{align}
	1&=\frac{\sqrt{\tilde{\epsilon}}}{\ell^2\sqrt{2\chi}} 
	\left[ 2 \sigma (\Sigma_1 -\Sigma_3) +  \Sigma_3 \right]. \\
	\intertext{multiplying Eq. (A16) by $\sqrt{2\chi}$, we get}
	\sqrt{2\chi} & = \frac{\sqrt{\tilde{\epsilon}}}{\ell^2} \left[ 2 \sigma (\Sigma_1 -\Sigma_3) +  \Sigma_3 \right]. \\ 
	\intertext{Squaring both sides and dividing by $2$, we find}
	\chi & = \frac{{\tilde{\epsilon}}}{2 \ell^4} \left[ 2 \sigma (\Delta \Sigma) +  \Sigma_3 \right]^2 ,
\end{align}

\noindent
where we denoted $\Delta \Sigma= \Sigma_1-\Sigma_3$. From Eqs. (A14) we also get

\begin{align}
	z_1+ \bar{z}_1 &=  \frac{1}{2\sqrt{\tilde{\epsilon}}\sigma}.\\ 
	\intertext{Squaring both sides yields}
	z_1^2+ \bar{z}_1^2 +2 z_1 \bar{z}_1 &=  \frac{1}{4\tilde{\epsilon} \sigma^2},\\ 
	\intertext{which we can write explicitly using the definitions of $z_i$ and $\bar{z}_i$, as}
	2\left( \lambda+ \tilde{u} c (1-\sigma) + \sqrt{2 \tilde{\epsilon} \chi \Sigma_1}\right) &=  \frac{1}{4\tilde{\epsilon} \sigma^2},\\
	\intertext{or}	
	\lambda+ \tilde{u} c (1-\sigma) + \sqrt{2 \tilde{\epsilon} \chi \Sigma_1} &=  \frac{1}{8\tilde{\epsilon} \sigma^2}.\\ 
	\intertext{Finally we can isolate $\lambda$}
	\lambda &=  \frac{1}{8\tilde{\epsilon} \sigma^2} -u c (1-\sigma) - \frac{\tilde{\epsilon}}{\ell^2} \sqrt{\Sigma_1} \left[2\sigma \Delta \Sigma +\Sigma_3\right].
\end{align}
Following the same process for $z_3$ we get
\begin{align}
	\lambda &=  \frac{1}{8\tilde{\epsilon} (1-2\sigma)^2} -2 u \tilde{u} \sigma - \frac{\tilde{\epsilon}}{\ell^2} \sqrt{\Sigma_3} \left[2\sigma \Delta \Sigma +\Sigma_3\right].
\end{align}

\noindent
Since these equations (A23) and (A24) express the same variable $\lambda$, we can equate both right hand sides to get an equation for $\sigma$

\begin{align} \label{eq: sigma eq full}
	\frac{1}{8\tilde{\epsilon} \sigma^2} - \frac{1}{8\tilde{\epsilon} (1-2\sigma)^2} - \tilde{u} c (1-3\sigma) - \frac{\tilde{\epsilon}}{\ell^2} \left(\sqrt{\Sigma_1}-\sqrt{\Sigma_3}\right)\left[2\sigma \left(\Sigma_1-\Sigma_3\right)+\Sigma_3\right] &=0.
\end{align}

\section{Metastable states}\label{app: metastable}

Since Eq.(37) in the main text may have more than one solution, stability classification is needed. Being an effective equation, stability analysis was done directly from this equation, using standard techniques borrowed from catastrophe theory, as explained in the the main text. 
We thus plot the same graphs as in Figs. 1-3 of the main text, just this time not only with the most stable solution, but rather with the unstable (green dash-dotted) and meta-stable (yellow dashed) curves as well.

\begin{figure}
	\includegraphics[width=0.95\textwidth]{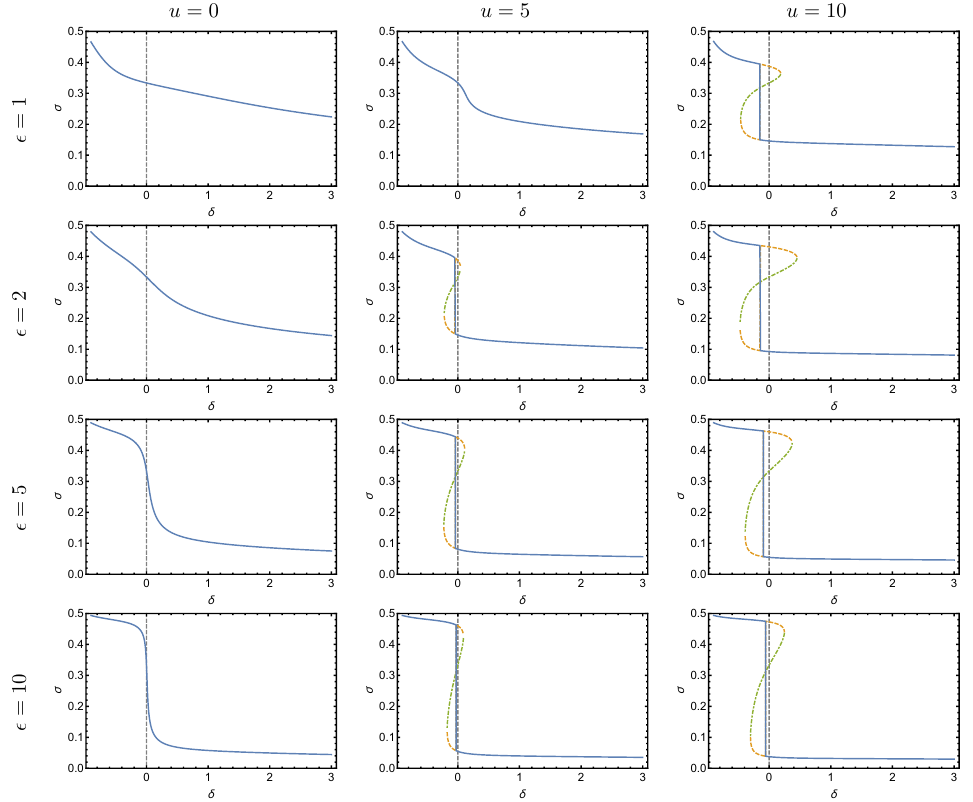}	
	\caption{$\sigma$ vs $\delta$, for different values of $\epsilon$ (rows) and $u$ (columns), the vertical gray dashed line marks the case of a sphere. Notice that since we assume a cylindrical symmetry in our analysis, even in a perfect sphere case we get a distinct stable state. The results at $\delta =0$ should be understood as the limit $\delta \rightarrow 0$, i.e. the limit of an infinitesimally broken symmetry. The region between this (gray) line and the vertical drop is the region where entropy prevails over energy for the oblate spheroid. 
		\label{fig: app_s_d}}
\end{figure}

\begin{figure}
	\includegraphics[width=0.95\textwidth]{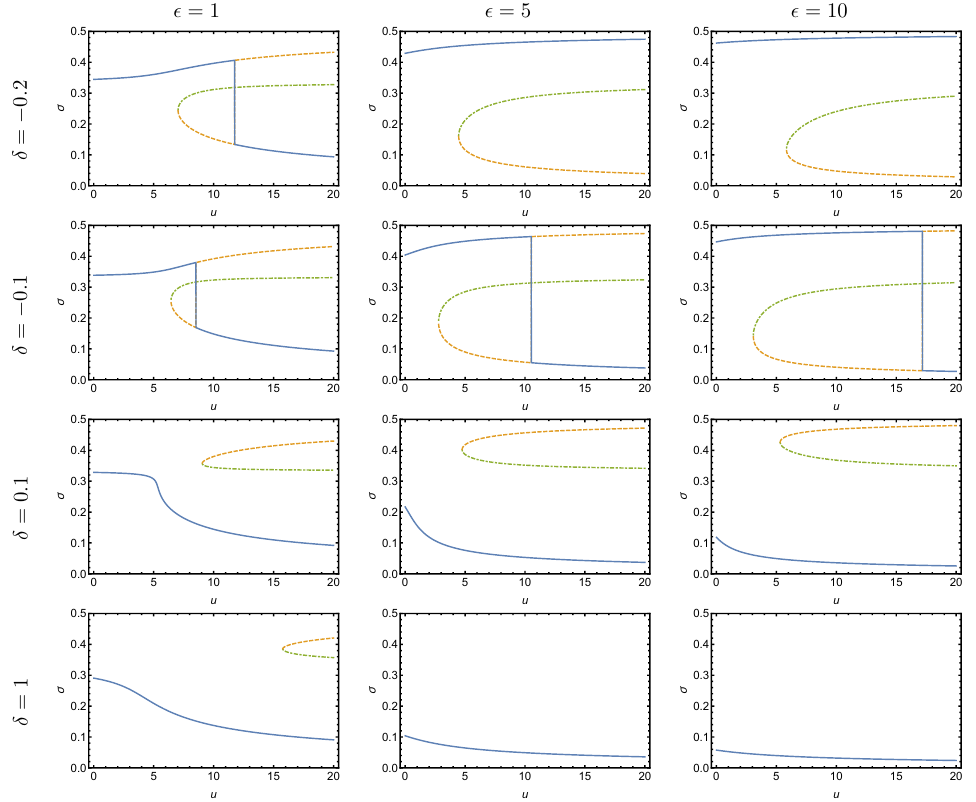}	
	\caption{$\sigma$ vs $u$, for different values of $\delta$ (rows) and $\epsilon$ (columns).
		\label{fig: app_s_u}}
\end{figure}

\begin{figure}
	\includegraphics[width=0.95\textwidth]{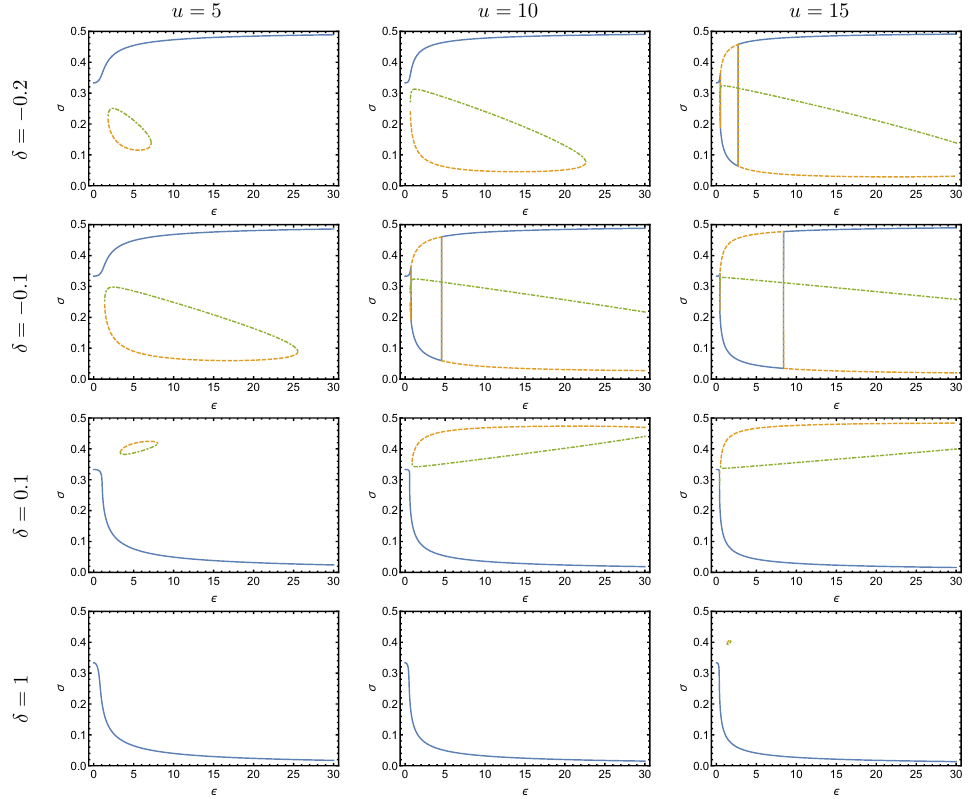}	
	\caption{$\sigma$ vs $\epsilon$, for different values of $\delta$ (rows) and $u$ (columns).
		\label{fig: app_s_e}}	
\end{figure}

Below, in Fig. \ref{fig: app_s_d} we present $\sigma$ plotted vs the flattening $\delta$ for various values of $u$ and $\epsilon$. Most notably we see that whenever the ground state appears to abruptly change, we find that in a neighbourhood aroud it there are $3$ solutions rather than a single one. An unstable solution always appears near $\sigma \sim \frac{1}{3}$. Most importantly we see that the existence region of the metastable states gets larger with the self-avoidance $u$.  Similar results can be seen in Figs. \ref{fig: app_s_u} and \ref{fig: app_s_e}.

\blu{Additionally, the Mathematica notebook "3D.nb" attached to this supplemental material shows a 3D, interactive, visualization of $\sigma$, near the the phase transition, as an additional way to view the complexity of the critical behavior of this system. $\sigma$ is plotted in the range $-0.1 \leq \delta \leq 0.1$ and $0 \leq u \leq 15$, for $\epsilon =5$. }

\section{Random filament} \label{app: random coil}

In this appendix we calculate the orientation order parameter $\sigma_{\rm rand}$ that describes 
a completely random filament over the surface of the ellipsoid.
Calculation of the random filament is done first by describing the configuration of an ellipsoid, using polar coordinates $\theta$ and $\phi$ as follows

\begin{align}
	\mathbf{f}\left(\theta,\phi\right) = \left\{\cos \phi \sin \theta, \sin\phi \sin\theta, \left(1+\delta\right)\cos\theta \right\} ,
\end{align}

\noindent
where $\delta$ is the flattening. Note that the axes scaling is constant, and does not change with $\delta$ as we eventually average over the area, and it does not play a role. 

We can now define a locally flat frame $\mathbf{t}_\theta = \frac{\pd_\theta \mathbf{f}}{\sqrt{\pd_\theta \mathbf{f}\cdot \pd_\theta \mathbf{f}}}$, $\mathbf{t}_\phi = \frac{\pd_\phi \mathbf{f}}{\sqrt{\pd_\phi \mathbf{f} \cdot \pd_\phi \mathbf{f}}}$ ,$\hat{\mathbf{n}} = \frac{\mathbf{t}_\theta \times \mathbf{f}_\phi}{\left|\mathbf{t}_\theta \times \mathbf{t}_\phi \right|}$. Where $\mathbf{t}_\theta$ and $\mathbf{t}_\phi$ are the surface tangents and $\hat{\mathbf{n}}$ is the normal.  A "2D" unit vector locally given by $r=\left(\cos\psi, \sin \psi\right)$, where $\psi$ is the angle relative to $\mathbf{t}_\theta$, has the 3D form

\begin{align}
	\mathbf{r}_{3D} =\cos\psi\mathbf{t}_\theta +\sin\psi\mathbf{t}_\phi.
\end{align}

\noindent
The orientation tensor $\sigma^{\mu \nu}(\psi, \theta,\phi)$ is given by $\sigma^{\mu\nu} = r_{3D}^\mu r_{3D}^\nu$.  The global, averaged tensor is then given by

\begin{align}
	\sigma_{\rm glob}^{\mu\nu} =
	\frac{\displaystyle \int {\left(\int  r_{3D}^\mu r_{3D}^\nu \dt \psi\right) \sqrt{\sin^2\theta\left(\cos^2\theta \left(1+\delta\right)^2\sin^2\theta\right)}\dt\theta\dt\phi}}
	{2\pi \displaystyle \int \sqrt{\sin^2\theta\left(\cos^2\theta \left(1+\delta\right)^2\sin^2\theta\right)}\dt\theta\dt\phi }.
\end{align}

\noindent
Comparing the result to the form $\left(\begin{array}{ccc}
	\sigma_{\rm rand} & 0 & 0 \\
	0 & \sigma_{\rm rand} & 0 \\
	0 & & 1- 2\sigma_{\rm rand}
\end{array}\right)$, we find

\begin{align}
	\sigma_{\rm rand} = \frac{\sqrt{\delta\left(2+\delta\right)}\left[\delta\left(2+\delta\right) - 1\right] + \left(1 + \delta\right)^4 \arccos\left(\frac{1}{1+\delta}\right)}{4 \delta \left(2 + \delta \right)\left[\sqrt{\delta \left(2 + \delta\right)} + \left(1 + \delta\right)^2 \arccos\left( \frac{1}{1+\delta} \right) \right]}.
\end{align}

\end{document}